\documentclass[%
 reprint,
 amsmath,amssymb,amsfonts,
 aps,
 prl,
 twocolumn,
 superscriptaddress,
 showpacs
]{revtex4-2}

\usepackage{graphicx}
\usepackage{dcolumn}
\usepackage{mathtools}
\usepackage{pgfplots}
\usepackage{tikz}
\usepackage{xcolor}

\usepgfplotslibrary{groupplots}
\pgfplotsset{every tick label/.append style={font=\tiny}}
\pgfplotsset{compat=1.18}

\definecolor{apsblue}{RGB}{16, 38, 148}
\usepackage[colorlinks=true, allcolors=apsblue,bookmarks=false]{hyperref}

\newcommand\br{ {\bf r}}
\newcommand\bfm{{\bf m}}

\newcommand\bu{ {\bf u}}

\newcommand\bF{{\bf F}}
\newcommand\U{U_{\rm int}}

\newcommand\bJ{{\bf J}}
\newcommand\subA{{ \rm A}}
\newcommand\subP{{ \rm P}}
\newcommand\subIK{{ \rm IK}}
\newcommand\bsigma{{\boldsymbol{\sigma}}}

\newcommand{\eff}{\mathrm{eff}}
\renewcommand\vec[1]{{\bf #1}}

\begin{document}

\preprint{APS/123-QED}
\title{{Revisiting the Ratchet Principle: When Hidden Conservation Laws Prevent Directed Currents in Stochastic Systems}}

\author{Jessica Metzger}
\affiliation{Department of Physics, Massachusetts Institute of Technology, Cambridge, Massachusetts 02139, USA}
\author{Sunghan Ro}
\affiliation{Department of Physics, Harvard University, Cambridge, Massachusetts 02138, USA}
\affiliation{Department of Physics, Massachusetts Institute of Technology, Cambridge, Massachusetts 02139, USA}
\author{Julien Tailleur}
\affiliation{Department of Physics, Massachusetts Institute of Technology, Cambridge, Massachusetts 02139, USA}

\date{\today}

\begin{abstract}
The ``ratchet principle", which states that non-equilibrium systems violating parity symmetry generically exhibit steady-state currents, is one of the few generic results outside thermal equilibrium. We study exceptions to this principle observed in active and passive systems with spatially varying fluctuation sources. For dilute systems, we show that a hidden time-reversal symmetry prevents the emergence of ratchet currents. At higher densities, pairwise forces break this symmetry but an emergent conservation law for the momentum field may nevertheless prevent steady currents. We show how the presence of this conservation law can be tested analytically and characterize the onset of ratchet currents in its absence. Our results show that the ratchet principle should be amended to preclude parity symmetry, time-reversal symmetry, and bulk momentum conservation.
\end{abstract}

\maketitle

Curie's Principle asserts that any {symmetry breaking at the microscopic scale must generically have observable consequences~\cite{curie_sur_1894}. An important illustration of this is} the `Ratchet Principle'~\cite{ajdari_mouvement_1992,magnasco_forced_1993,magnasco_molecular_1994,rousselet_directional_1994,ajdari_rectified_1994,parrondo1996criticism,reimann_brownian_2002,galajda_wall_2007,hanggi_artificial_2009,di2010bacterial,sokolov2010swimming,angelani_active_2011,reichhardt_ratchet_2017,obyrne_time_2022,rein_force-free_2023}, which has become a cornerstone of non-equilibrium statistical mechanics. This principle states that a system which violates {properly defined} time-reversal and parity symmetries should generically display a nonvanishing steady-state {directed current~\cite{denisov_tunable_2014}. Such currents, to which we refer as `ratchet currents', trace back to the work of Smoluchowski~\cite{smoluchowski1912experimental} and Feynman~\cite{feynman_feynman_1963} on the interplay between fluctuations and asymmetric potentials. Since then, ratchet currents have been found at all scales in nature, from the motion of motor proteins~}\cite{julicher_modeling_1997,frey2005brownian,campas2006collective} {to the powering of asymmetric gears by bacterial suspensions~}\cite{di2010bacterial,sokolov2010swimming,anand_transport_2024}. {The scope of the ratchet principle has significantly broadened over time and it has been shown that diverse forms of parity-symmetry breaking, beyond that due to external potentials, may induce directed `ratchet' currents}~\cite{kidoaki_rectified_2013,caballero_ratchetaxis_2015,rein_force-free_2023,borsley_molecular_2024}.

\begin{figure}
    \begin{tikzpicture}
        \def\x{-3.27}
        \path (-0,0) node {\includegraphics[width=\columnwidth]{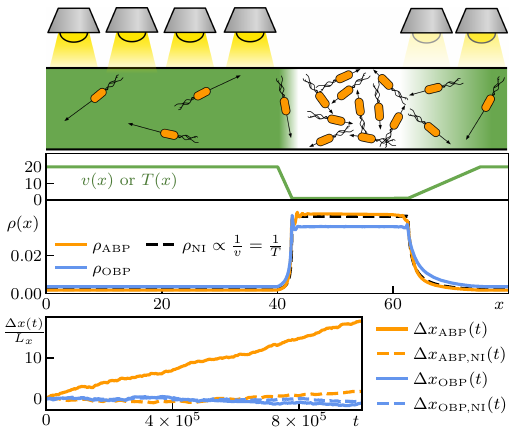}};
        \draw (\x,2.26) node[fill=white, fill opacity=1, draw=black, text opacity=1, inner sep=1.5pt] {(a)};
        \draw (\x,0.83) node[fill=white, fill opacity=1, draw=black, text opacity=1, inner sep=1.5pt] {(b)};
        \draw (\x-.01,0.05) node[fill=white, fill opacity=1, draw=black, text opacity=1, inner sep=1.5pt] {(c)};
        \draw (\x,-1.92) node[fill=white, fill opacity=1, draw=black, text opacity=1, inner sep=1.5pt] {(d)};
    \end{tikzpicture}
    \caption{{\bf (a)}: Light-activated particles, like bacteria whose flagella rotors are controlled by proteorhodospin~\cite{arlt_painting_2018,frangipane_dynamic_2018}, accumulate in low-{speed} regions. {\bf (b)}: {Speed} and temperature fields. 
    {\bf (c)}: Steady-state density of ABPs and {OBPs} interacting via soft repulsive forces {with} the fluctuation {sources} shown in (b), compared {to} the non-interacting {case}.
    \textbf{(d)}: Interactions induce a steady {directed} current for ABPs, but not for {OBPs}. All numerics are detailed in End Matter.}
    \label{fig:ratchet-exception-vis}
\end{figure}

However, there are exceptions to the ratchet principle{:~Some systems violate time-reversal and parity symmetries but lack steady currents, which calls for an explanation. For instance, some systems admit additional symmetries such that trajectories leading to opposite values of the current are equiprobable, which makes the average current vanish~\cite{denisov_tunable_2014}. Such hidden symmetries have received a lot of attention for ratchets induced by external potentials~}\cite{flach_directed_2000,denisov_tunable_2014,cubero_hidden_2016}{. Another case where directed currents fail to emerge is that of asymmetric fluctuation landscapes, in the absence of external potentials~}\cite{van_kampen_diffusion_1988,schnitzer_theory_1993,cates_when_2013,martin_aoup_2021}{.
Indeed, noninteracting overdamped Brownian particles (OBPs) in inhomogenous temperature fields admit flux-free steady states~}\cite{van_kampen_diffusion_1988,van_kampen_relative_1988}{.
In active matter, where ratchets have attracted a lot of attention~}\cite{fiasconaro_active_2008,angelani_active_2011,pototsky_rectification_2013,ghosh2013self,lopatina_self-driven_2013,ai_rectification_2013,reichhardt_active_2013,koumakis_directed_2014,yariv_ratcheting_2014,   bijnens_pushing_2021,martin_aoup_2021,derivaux_rectification_2022,zhen_optimal_2022,khatri_inertial_2023,muhsin_inertial_2023,rojasvega_mixtures_2023,ryabov_mechanochemical_2023,schimming_active_2024,anand_transport_2024,wang_spontaneous_2024}{, spatially varying propulsion speeds do not induce directed currents for noninteracting run-and-tumble particles (RTPs)~\cite{schnitzer_theory_1993} or active Brownian particles (ABPs)~\cite{cates_when_2013}. It was however shown  numerically that the ABP current-free state is fragile: repulsive interparticle forces suffice to restore a steady current in the presence of asymmetric activity landscape~\cite{stenhammar_light-induced_2016}---a case that has not been explored for OBPs or RTPs. Currently, what prevents the emergence of currents in asymmetric fluctuation landscapes is unknown. Moreover, how interactions restore directed currents in this context is a particularly exciting open  question, given the existing interest in interaction-induced ratchets~}\cite{reimann_brownian_2002,van_der_meer_spontaneous_2004,liebchen_interaction-induced_2012,liebchen_interaction_2015}{. All in all, these issues call into question the scope of the ratchet principle.}

In this Letter, we build a theoretical framework to account for the emergence of {interaction-induced directed} currents in systems with inhomogenous fluctuation sources, which leads to the conclusion that the ratchet principle must be amended. We consider {$N$ particles} in $d$ space dimensions {evolving with}
\begin{align}
        \dot{\br}_i = - \sum_j \nabla \U(\br_i-\br_j) + \mathbf{F}_i(\br_i,t)\;,
    \label{eq:langevin}
\end{align}
where $\U$ is an interaction potential, the particle mobility is set to 1, and $\mathbf{F}_i$ is a fluctuating force. In the active case, $\mathbf{F}_i(\br_i,t) = v(\br_i) \bu_i(t) $, where the orientation $\bu_i$ either tumbles at constant rate $\tau^{-1}$ for RTPs, or undergoes rotational diffusion {with rotational} diffusivity $1/[(d-1)\tau]$ for ABPs. For {OBPs}~\cite{van_kampen_relative_1988},  Eq.~\eqref{eq:langevin} is an It\={o}-Langevin equation and $\mathbf{F}_i(\br_i,t) = \sqrt{2 T(\br_i(t))}\boldsymbol{\eta}_i(t)$, with $\boldsymbol{\eta}_i$ a unit-variance centered Gaussian white noise and $k_B=1$.
{Throughout our work, we focus on the setup depicted in Fig.~\ref{fig:ratchet-exception-vis} in which fluctuation sources are modulated along $\hat x$ and we study the condition under which a non-vanishing steady current $\bJ=J \hat x$ emerges.}
First, we consider the noninteracting case where we show that the lack of directed current stems from a hidden detailed balance with respect to a non-Boltzmann steady state: the non-interacting dynamics are, in fact, time reversible.
Both passive and active systems become irreversible once interactions are introduced, but only the active case leads to {directed currents}, as shown numerically in Fig.~\ref{fig:ratchet-exception-vis}(d). The {OBP} ratchet exception is thus robust to interactions.
We show that this robustness {stems from} an {effective} conservation law for the OBP momentum field, which forbids the momentum sources required to power directed currents. On the contrary, the dynamics of interacting ABPs and RTPs with non-uniform self-propulsion speeds allow for net momentum sources that violate this effective conservation law. In addition to characterizing these momentum sources numerically, we develop a theoretical framework that accounts for them analytically, which we illustrate on RTPs in {1d}. Our formalism allows us to predict quantitatively the steady-state current in the high-density limit and to confirm from first principles the lack of current for {OBPs}. All in all, our results thus show that the ratchet principle must be amended to preclude bulk momentum conservation. Detailed derivations, together with generalizations to other models including Active Ornstein Uhlenbeck particles{---which do not display directed currents, event in the presence of interactions---}are provided in a companion article~\cite{metzger_exceptions_companion}.

\textit{{Hidden time-reversal symmetry.}} Let us first show that the lack of current in the non-interacting case stems from a hidden {time-reversal symmetry}.

In the active case, the dynamics lead to a current-free steady state with a density {field} $\rho(\br){\equiv \langle\sum_i \delta(\br-\br_i)\rangle}\propto v^{-1}(\br)$~\cite{schnitzer_theory_1993,cates_when_2013}. For parity-breaking $v(\br)$, this non-thermal {yet current-free} dynamics seems to escape the ratchet principle: The entropy-production rate of, say, a run-and-tumble particle in the full $(\br,\bu)$ space is indeed positive~\cite{razin2020entropy,cocconi2020entropy,frydel2022intuitive,paoluzzi2024entropy}. {The lack of current in this parity-breaking irreversible dynamics thus appears as violating the ratchet principle.}

The paradox is resolved by noticing that currents are determined by changes in positions $\br(t)$. We therefore coarse-grain out the variable $\bu(t)$ and {compute} the entropy production rate in $\br$-space{, defined as}:
\begin{equation}
    \sigma = \lim_{t_{\rm f} \to \infty} \frac 1 {t_{\rm f}} \log \frac{\mathbb{P}[\br^R(t)]}{\mathbb{P}[\br(t)]}\;,
    \label{eq:trs-rtp}
\end{equation}
where $\br^R(t)=\br(t_{\rm f}-t)$ is the time-reversal of a trajectory of duration $t_{\rm f}$, and $\mathbb{P}$ the probability density of trajectories. {We note that $\sigma$ does not measure the rate of change of the total thermodynamic entropy of the system. Instead, it is an information-theoretic measure which tells us whether particle trajectories $\br(t)$ played forward and backward can be statistically distinguished~\cite{o2022time,fodor2022irreversibility}: it is a direct measure of irreversibility in position space. Thermodynamic and information-theoretic entropy-production rates may coincide~\cite{seifert_entropy_2005}, but they often do not~\cite{celani_anomalous_2012,fodor2022irreversibility}, especially in active matter~\cite{shankar2018hidden,o2022time}. For non-interacting RTPs and ABPs in activity landscapes,} direct computations~\cite{metzger_exceptions_companion} show that $\sigma=0$ in the absence of interactions. Their dynamics thus satisfy detailed balance with respect to their {non-thermal steady states} in position space: {the lack of currents follows from this hidden time-reversal symmetry and obeys the ratchet principle~\cite{denisov_tunable_2014}.}
\begin{figure}
    \begin{tikzpicture}
        \path (0,0) node {\includegraphics{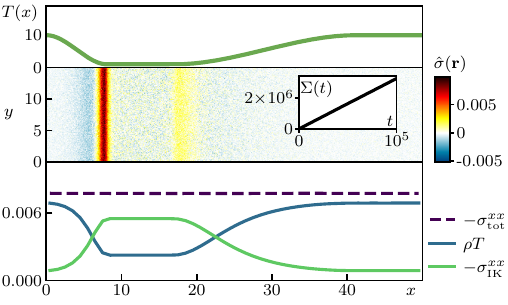}};
        \def\x{-3.25}
        \draw(\x,2.26) node[fill=white, draw=black, fill opacity=1, text opacity=1, inner sep=1.5pt] {(a)};
        \draw(\x,1.22) node[fill=white, draw=black, fill opacity=1, text opacity=1, inner sep=1.5pt] {(b)};
        \draw(\x,-.38) node[fill=white, draw=black, fill opacity=1, text opacity=1, inner sep=1.5pt] {(c)};
    \end{tikzpicture}
    \caption{{Entropy production rate and stress measured in simulations of OBPs in the temperature landscape shown in panel {\bf (a)}. {\bf (b)} Heat map of the entropy production rate density field $\sigma(\br)=\langle \hat{\sigma}(\br)\rangle$. Inset: Net entropy production up to time $t$. {\bf (c)} Plot of the $(x,x)$ component of the Irving-Kirkwood stress tensor and of the ideal gas pressure. Their sum is uniform, which prevents the emergence of directed currents.}}
    \label{fig:EOSpassive}
\end{figure}

For non-interacting {OBPs} in inhomogenous temperature fields, {$\sigma=0$} despite the non-Boltzmann steady state {$\rho(\br)\propto 1/T(\br)$}~\footnote{{Note that the large damping limit of underdamped Brownian particles has a non-vanishing \textit{thermodynamic} entropy production rate, showing the disconnection between thermodynamics and information-theoretic measures in this case~\cite{celani_anomalous_2012,metzger_exceptions_companion}.}}. The lack of {directed current in asymmetric $T(\br)$  thus also stems from a hidden time-reversal symmetry in position space.} This highlights the need to break detailed balance, not just depart from thermal equilibrium, to observe {directed} currents~\cite{denisov_tunable_2014}.

{Let us now turn to interacting} systems ($\U\neq 0$). {In the active case, the dynamics are irreversible ($\sigma>0$)}, even with homogenous activity~\cite{fodor_how_2016,o2022time}, a result that directly extends to activity landscapes. The emergence of currents in {this case~\cite{stenhammar_light-induced_2016} is thus compatible with the ratchet principle. In the passive case, $\sigma$ is also positive in the presence of interactions. Direct algebra~\cite{metzger_exceptions_companion} shows that $\sigma(t) = \langle \int d{\bf r} \hat\sigma(\br,t)\rangle$, where we use Stratonovich convention and}
\begin{equation}
    {\hat \sigma(\br)\equiv -\sum_{n,m=1}^N \frac{ \dot{\br}_n \cdot \nabla U_{\text{int}}(\br_n-\br_m)}{T(\br_n)} \delta(\br-\br_n)\;.}
    \label{eq:epr-OBP}
\end{equation}
{Figure~\ref{fig:EOSpassive}(b) shows that $\Sigma(t)=\int_0^t \sigma(s)ds>0$, despite the lack of directed currents in Fig.~\ref{fig:ratchet-exception-vis}(d), which violates the ratchet principle. We note that  irreversibility sometimes vanishes upon coarse-graining~\cite{o2022time}, leading to effective large-scale equilibrium physics. This is not the case for the interacting systems considered here, and it  would still not explain the exact vanishing of the current at the microscopic scale. An explanation of the lack of directed current for interacting OBPs in asymmetric $T(\br)$ is thus required.} Let us now show that it stems from the lack of net momentum sources in the system.

\textit{Momentum conservation prevents {directed} currents.}  We first recall the connection between stress tensor and bulk momentum conservation.  In classical mechanics, Newton's {second} law implies that the momentum field of {particles with mass $m_\mathrm{p}$}, $\vec p(\br,t)=\langle\sum_i {m_\mathrm{p}} \vec v_i \delta[\br-\br_i(t)]\rangle$, evolves as~\footnote{{The divergence operator $\nabla \cdot$ contracts with the first index of the following tensor.}} $\dot {\vec p} = - \nabla \cdot \langle {m_\mathrm{p}} \sum_i \vec v_i \otimes \vec v_i \delta[\vec r-\vec r_i(t)]\rangle + \vec f_\rho(\vec r)$, where $\vec f_\rho(\vec r)\equiv \langle \sum_i \vec f_i \delta[\vec r-\vec r_i(t)] \rangle$ is the local force density. {Irving and Kirkwood have shown how momentum conservation implies that the force density can be written as $\vec f_\rho=\nabla\cdot \bsigma_{\subIK}$, where $\sigma_{\rm IK}^{\alpha\beta}$ measures the flux in the direction $\hat \alpha$ of momentum along $\hat \beta$ due to interparticle forces}~\cite{irving1950statistical}. {(The derivation of $\bsigma_{\rm IK}$ is detailed in End Matter for completeness.)} The evolution of $\vec p$ then reads $\dot {\vec p} = \nabla \cdot\bsigma$, where $\bsigma$ is the total stress tensor.  In the presence of damping due to a solvent, the dynamics become $\dot {\vec p} = \nabla \cdot \bsigma - \gamma \vec p$. Since the particle current is $\bJ=\vec p/{m_\mathrm{p}}$, the momentum-conserving nature of the forces can be read in the steady-state relation $\bJ = \mu \nabla \cdot \bsigma$, where $\mu=(\gamma {m_\mathrm{p}})^{-1}$ is the particle mobility. {Immediately, this leads to $\int d \br \bJ=0$ and the lack of directed current.}

Let us now show that {OBPs} in inhomogenous temperature fields satisfy such {a conservation law for momentum}. In $d$ dimensions, It\=o's formula shows that Eq.~\eqref{eq:langevin} makes $\rho$ evolve as ${\dot\rho}=-\nabla \cdot \bJ$, with
\begin{equation}
 \bJ = \nabla \cdot \bsigma_\subP,\quad\text{where}\quad    \bsigma_\subP(\br) = \bsigma_{\subIK}(\br) - \rho(\br) T(\br) \mathbb{I}_{d}\;.\label{eq:eos-OBP}
\end{equation}
Despite the non-uniform temperature, the system admits a well-defined
stress tensor $\bsigma_\subP$ and {$\int d \br \bJ=0$: Temperature inhomogeneities thus} cannot act as {net} momentum sources to drive a {steady directed current, as confirmed by the numerical simulations shown} in Fig.~\ref{fig:EOSpassive}{(c)}.
\begin{figure}
    \begin{tikzpicture}
        \path (0,0) node {\includegraphics[width=.95\columnwidth]{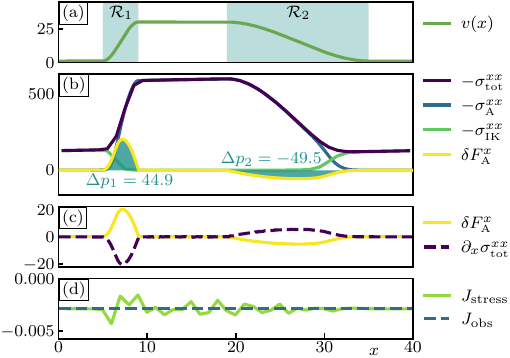}};
    \end{tikzpicture}
    \caption{Simulations of interacting ABPs in the activity landscape shown in panel {\bf (a)}. {\bf (b)} As in the passive case (Fig.~\ref{fig:EOSpassive}), the contributions to the total stress---$\bsigma_{\subA}$  and $\bsigma_{\subIK}$---vary throughout space. Here, however, the total stress {$\bsigma_{\rm tot}$} also varies due to the momentum source $\delta F_\subA$. The {exchanged} momentum $\Delta p_1$ in region $\mathcal{R}_1$ and $\Delta p_2$ in $\mathcal{R}_2$ {do not cancel out}, leading to the {directed} current $\langle J\rangle$ shown in panel (c). Outside $\mathcal{R}_1$ and $\mathcal{R}_2$, the current is supported by non-uniform density profiles, which lead to a non-uniform stress.  {\bf (c)} {The momentum} source $\delta F_A$ and the variation of the total stress add up to produce a uniform current,  $J_{\rm stress} = \partial_x \sigma_{\rm tot}^{xx} + \delta F_\subA^x$. It matches the current $J_{\rm sim}$ measured directly by tracking particle displacements.}
    \label{fig:momentum-source-abp}
\end{figure}

\textit{Momentum sources for ABPs and RTPs.}  The pressure of ABPs and RTPs {with constant self-propulsion speed $v_0$ and pairwise interaction forces admits} an equation of state~\cite{solon_pressure_2015,fily_mechanical_2018,granek_colloquium_2024}. {This stems from} their active-force density {being given by} the divergence of {an active stress tensor $\bsigma_{\rm A}$, such that $\sigma_{\rm
  A}^{\alpha\beta}$ measures the flux in direction $\hat \alpha$ of
the average momentum in direction $\hat \beta$ that the active force
extracts from the substrate~\cite{fily_mechanical_2018}}. As shown in~\cite{wysocki_interacting_2022}, this effective momentum
conservation is violated in the presence of an activity landscape
$v(\br)$. Let us now show how the corresponding momentum sources can
power interaction-induced {directed currents.}
 
{Using the chain rule and Eq.~\eqref{eq:langevin}, the exact dynamics of the density field reads}~\cite{solon_pressure_2015-1,solon2018generalized,solon2018generalizedNJP,speck2021coexistence,wysocki_interacting_2022,omar2023mechanical}:
\begin{equation}
    {\dot{\rho}} = -\nabla \cdot \left[v(\br) \bfm - \int d \br' \rho_2(\br,\br') \nabla \U(\br-\br') \right]\;,\label{eq:abp-rtp-rhodot}
\end{equation}
where we have introduced {the two-point density} $\rho_2(\br,\br') =
\langle \sum_{i,j} \delta(\br-\br_i) \delta(\br'-\br_j)\rangle$ {and the orientation field $\bfm(\br)=\langle\sum_i \bu_i
\delta(\br-\br_i)\rangle$. In turn, It\=o's formula shows $\bfm$ to evolve
as}
\begin{equation}
  {\dot \bfm = -\nabla \cdot \bJ_m - \frac \bfm \tau, \quad \bJ_m \equiv\langle \sum_i \dot \br_i \otimes \bu_i \delta(\br-\br_i)\rangle }\label{eq:abp-rtp-mdot},
\end{equation}
{where $\bJ_m$ is a tensor measuring the flux of polar order.}

{We then introduce $\bsigma_{\subA}(\br)=\tau \bJ_m v(\br)$, which coincides with the active stress tensor of ABPs and RTPs with  constant speeds $v_0=v(\br)$. In the steady state, $\dot{\bfm}=0$ and Eq.~\eqref{eq:abp-rtp-mdot} yields}
\begin{equation}
    {v \bfm = -\nabla\cdot \left[\tau \bJ_m v \right] + \tau \bJ_m \cdot \nabla v \equiv \nabla \cdot \bsigma_{\subA} + \delta \bF_\subA\;,}\label{eq:abp-rtp-m-ss}
\end{equation}
{where we have introduced the momentum source $\delta \bF_\subA(\br)$} that arises from the nonuniformity of
$v(\br)$:
\begin{equation}
{\delta \bF_\subA(\br) \equiv \tau \bJ_m(\br) \cdot \nabla v(\br)\;.}
\label{eq:abp-rtp-eos-violation}
\end{equation}
{As in the passive case, the interaction term in Eq.~\eqref{eq:abp-rtp-rhodot}
can be written as the divergence of the Irving-Kirkwood stress
$\bsigma_\subIK$, leading to} $\dot{\rho}=-\nabla \cdot \bJ$ with
\begin{equation}
    \bJ =  \delta \bF_{\subA}+\nabla \cdot \bsigma_{\rm tot} ,\quad\text{and}\quad\bsigma_{\rm tot}=\bsigma_\subIK + \bsigma_{\subA}\,.
    \label{eq:momentumsource}
\end{equation}

{Equations~\eqref{eq:abp-rtp-eos-violation} and~\eqref{eq:momentumsource} show how the inhomogenous activity ($\nabla v\neq 0$) induces a} momentum source $\delta\bF_\subA$. In Fig.~\ref{fig:momentum-source-abp}, {we show numerically that the total} injection of momentum {by $\delta\bF_\subA$ for an asymmetric profile $v(x)$  quantitatively accounts for the measured steady current}. In contrast to {OBPs} in a temperature landscape, {net} momentum sources in ABPs and RTPs thus allow for the emergence of {directed} currents in the presence of spatially varying speeds and pairwise forces: {time-reversal symmetry} and momentum conservation are both violated, allowing {our amended} ratchet principle to apply.

The results so far show on general grounds that the ratchet principle should be amended to require the existence of net momentum sources. In the above, we numerically demonstrated their existence for ABPs (Fig.~\ref{fig:momentum-source-abp}). We now present a theoretical framework to predict quantitatively the  emergence of interaction-induced {directed} currents. We {consider RTPs in 1d} and show how {our} formalism correctly discriminates with the case of {OBPs}.

\begin{figure}
\begin{tikzpicture}
    \path (0,0) node {\includegraphics{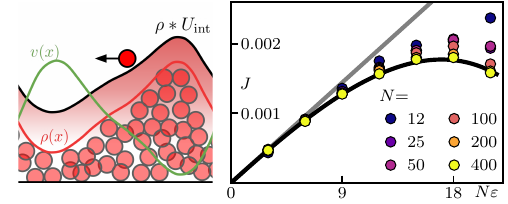}};
    \def\y{1.47}
    \draw(-3.9,\y) node {(a)};
    \draw(-0.1,\y) node {(b)};
\end{tikzpicture}
    \caption{{\bf (a)} At the mean-field level, an RTP (bright red) moving in an activity landscape $v(x)$ (green) experiences an effective external potential $V_\eff = U_{\text{int}}\ast \rho$ induced by its neighbors (dim red). {\bf (b)} Comparison between the current measured in simulations (circles) and its prediction by the perturbative solution of Eqs.~\eqref{eq:CWr}-\eqref{eq:CWm}, as the scaled interaction strength $N\varepsilon$ is varied. The first order, $J_1$ (gray line), captures the slope of $J$ at $\varepsilon=0$ while the sixth order (black line) captures the non-monotonicity of $J$ as $N \varepsilon$ increases. Note that the radius of convergence of the expansion is limited by a jamming transition that occurs at larger values of $N\varepsilon$.}
    \label{fig:mean-field-schematic}   \label{fig:perturbative-ratchet}
\end{figure}

\textit{Interaction-induced current in {1d} RTPs.} We consider Eq.~\eqref{eq:langevin} {in one dimension, setting} the particle orientations to $\bu_i=s_i \hat x$ with $s_i=\pm 1$. {Equations}~\eqref{eq:abp-rtp-rhodot} and~\eqref{eq:abp-rtp-mdot} then reduce to
\begin{align}
    \dot\rho &= -\partial_x \Big[v m - \int dx' \rho_2(x,x') \U'(x-x')\Big]\label{eq:dynr1}\\
    \dot m &= -\partial_x \Big[v \rho - \int dx' m_2(x,x') \U'(x-x')\Big]- \frac m \tau\;,\label{eq:dynm1}
\end{align}
with $m_2(x,x') = \langle\sum_{i,j} s_i \delta(x-x_i) \delta(x'-x_j)\rangle$. To proceed, we consider the simplest factorization~\footnote{{In 1d, this mean-field approximation is expected to work well perturbatively in the interaction potential.}}, $\rho_2(x,x') \approx \rho(x) \rho(x')$ and $m_2(x,x') \approx m(x) \rho(x')$, which leads to:
\begin{align}
        {\dot \rho} &= -\partial_x \big[v m - \rho \partial_x V_{\text{eff}}\big]\label{eq:CWr}\\
        {\dot m} &= -\partial_x \big[v \rho - m \partial_x V_{\text{eff}}\big] - m        /\tau\;,\label{eq:CWm}
\end{align}
where we have introduced the effective potential $V_{\text{eff}}(x) \equiv (\U \ast \rho)(x) = \int dx' \U(x-x') \rho(x')$. In this Curie-Weiss mean-field picture, each particle evolves in an effective potential that results from the forces exerted by its neighbors, as depicted in Fig.~\ref{fig:mean-field-schematic}a. Since $V_{\rm eff}$ depends on $\rho$, Eqs.~\eqref{eq:CWr}-\eqref{eq:CWm} form a non-linear system of equations that must be solved self-consistently. 

\begin{figure}
\begin{tikzpicture}
    \path (0,0) node {\includegraphics{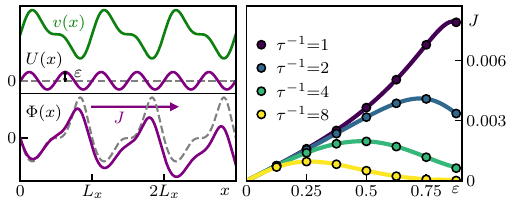}};
\def\x{-4.25}
\def\y{1.4}
    \draw (\x,\y) node {(a)};
    \draw (\x,-0.1) node {(b)};
    \draw (0.15,\y) node {(c)};
\end{tikzpicture}
    \caption{Non-interacting {1d} RTPs in {the} potential landscape $U(x)$ and activity landscape $v(x)$ depicted in {\bf (a)}. {\bf (b)} Pseudo potential $\Phi(x)$ in the presence (solid line) or in the absence (dashed line) of $U(x)$. An aperiodic $\Phi$ leads to a steady current $J$.  {\bf (c)} Comparison between the steady-state current measured in particle simulations (symbols) and its predictions from Eq.~\eqref{eq:exactRTP} (lines) for various {tumbling rates $\tau^{-1}$}. }
    \label{fig:rtp-u-v-exact-sol}
\end{figure}

Before doing so, it is instructive to consider Eqs.~\eqref{eq:CWr}-\eqref{eq:CWm} for an arbitrary external potential $V_{\rm eff}\to V$, a problem that can be solved exactly {using standard methods~\cite{tailleur_statistical_2008}. Rewriting}~\eqref{eq:CWr} as a continuity equation $\dot \rho=-\partial_x {J}$, {direct algebra, detailed in~\cite{metzger_exceptions_companion}, show the} particle current $J$ {to be given by:}
\begin{align}
    {J}&{\equiv \frac{\tilde F \rho - (\tilde T \rho)'}{\tilde\gamma},
    \qquad \tilde F= vv' \tau-V' - \frac{ (V')^2 v' \tau}{ v}}\label{eq:fpe-rtp-U-v1}\\
        {\tilde\gamma}&{=1- V'' \tau+\frac{v'V' \tau}{ v},
        \qquad \tilde T=\tau (v^2 - (V')^2).}   \label{eq:fpe-rtp-U-v}
\end{align}
The current-density relation~\eqref{eq:fpe-rtp-U-v1}-\eqref{eq:fpe-rtp-U-v} is equivalent to that of an overdamped passive particle with spatially varying friction $\tilde \gamma$, temperature $\tilde T$, and force field $\tilde F$ studied by van Kampen~\cite{van_kampen_relative_1988}. One {can then} solve for $\rho$ as~\footnote{We have fixed a typo in the {expression of $\rho(x)$} given in~\cite{van_kampen_relative_1988}.}:
\begin{equation}\label{eq:exactRTP}
    \rho(x)=\frac{\tilde T(0)}{\tilde T(x)}\rho_0 e^{-\Phi(x)}-J \int_0^x du \frac{\tilde\gamma(u)}{\tilde T(x)} e^{\Phi(u)-\Phi(x)}\;,
\end{equation}
{where we introduced the pseudo-potential}
\begin{equation}
{\Phi(x) \equiv -\int_0^x \frac{\tilde F(x')}{\tilde T(x')}dx'\;.}\label{eq:phi-def}
\end{equation}
This solution perfectly agrees with {the simulations} in
Fig.~\ref{fig:rtp-u-v-exact-sol}.  {Equation~\eqref{eq:exactRTP}
for $x=L$ shows $J\propto \rho_0[1-e^{-\Phi(L)}]$ so that a directed
current emerges iff} $\Phi(L)\neq 0$.

This result can now be used to account for the emergence of interaction-induced currents. In the mean-field picture, to leading order in the interaction strength, the steady-state density is $\rho(x)\simeq \kappa/ v(x)$, where $\kappa$ is a normalization constant, so that particles experience an effective potential $V_\eff \simeq  \kappa U_{\text{int}} \ast v^{-1}$. To leading order in $U_{\rm int}$, the condition for a non-vanishing current thus reads
\begin{align}
    \Phi(L) &= \frac{\kappa}{{\tau}} \int_0^L dx\int_0^L dx' \frac{U_{\text{int}}'(x-x')}{v(x') v(x)^2}  \neq 0\;.
    \label{eq:interaction-induced-ep}
\end{align}
{This stems directly from the definition of $\Phi$ in Eq.~\eqref{eq:phi-def}, upon the leading-order replacements $\tilde F(x)=-\rho_0 \ast \U$, $\rho_0=\kappa/v(x)$, and $\tilde T(x)= \tau v^2(x)$. While the numerator is antisymmetric in $x\leftrightarrow x'$, the denominator is not, hence leading to a non-vanishing current for a generic $v(x)$.} 
For a soft repulsive potential $\U$ of typical scale~$\varepsilon$, Eqs.~\eqref{eq:CWr}-\eqref{eq:CWm} can be solved perturbatively to arbitrary order in $\varepsilon$ using $\rho(x) = \sum_k \varepsilon^k \rho_k$ and $J = \sum_k \varepsilon^k J_k$. 
To first order, we find $J_1=-\kappa^2 {\tau} \Phi(L)$, where $\Phi(L)$ is given in {Eq.~}\eqref{eq:interaction-induced-ep}. {Figure~\ref{fig:mean-field-schematic} shows} that, in the large-density limit, this perturbative solution allows us to quantitatively account for the emergence of a {directed} current as the interaction strength increases. 

{Similar reasoning also predicts} the lack of current in {OBPs} to leading order in $\U$. Indeed, {Eqs.~\eqref{eq:exactRTP}-\eqref{eq:phi-def} also apply to OBPs in an external potential $V$, with $\tilde T=T(x)$, $\tilde \gamma=1$ and $\tilde F=-V'$~\cite{van_kampen_diffusion_1988}. To leading order in $\U$, using $\tilde F=-\partial_x [\U \ast \rho_0]$ with $\rho_0(x) =\kappa/T(x)$ leads to}
\begin{equation}
    \Phi(L) = \kappa \int_0^L dx \int_0^L dx' \frac{U_{\text{int}}'(x-x')}{T(x) T(x')} = 0\;,
\end{equation}
where we used that the denominator is symmetric in $x\leftrightarrow x'$, in contrast to {Eq.~}\eqref{eq:interaction-induced-ep}. 

\textit{Conclusion.} In this Letter, {we have shown how hidden time-reversal symmetries and momentum conservation laws prevent the emergence of steady currents in inhomogeneous fluctuation sources. 
For noninteracting RTPs and {OBPs}, the lack of currents indeed stems from a detailed balance in position space with respect to non-Boltzmann steady states.} 
{Time-reversal symmetry} is destroyed by interactions, but it only induces a current for RTPs and ABPs. 
OBPs are indeed protected from {directed currents because thermal fluctuations remain momentum-conserving in the presence of temperature gradients}.
The ratchet principle should thus be amended accordingly, to require not only the lack of time-reversal and parity symmetry{~\cite{denisov_tunable_2014}}, but also the existence of net momentum sources. 
We note that the violation of bulk-momentum conservation does not simply distinguish passive and active systems: in our companion publication~\cite{metzger_exceptions_companion}, we show that interacting active Ornstein Uhlenbeck particles, like {OBPs, remain free of directed current} in the presence of asymmetric activity landscapes. Beyond these general symmetry considerations, we have also shown how a mean-field theoretical framework allows predicting the emergence of interaction-induced directed currents, or the lack thereof.

Our work not only better delimits the scope of the ratchet principle, but also shows how activity, understood as the existence of net momentum sources, is crucial to power efficient microscopic motors, a question that has attracted a lot of attention recently~\cite{zakine2017stochastic,pietzonka2019autonomous,datta2022second}. {Extending our formalism to such microscopic machines appears  an exciting avenue. Furthermore, we have considered in this Letter the currents in the colloidal subsystem. In the presence of momentum-conserving fluids, it must be accompanied by counter-flows~\cite{malgaretti2017bistability} whose properties remain to be studied.}

\textit{Acknowledgements.} We thank Hughes Chate, {Cory Hargus, }Yariv Kafri, Jeremy O'Byrne, and Vivien Lecomte for insightful discussions.

~\newpage{}

\appendix

\section{End Matter}
All simulations in this Letter use an Euler integration scheme, with time in units of the persistence time and space in units of the interaction radius, when applicable. The data shown in Fig.~\ref{fig:ratchet-exception-vis} come from particle-based simulations of {2d} ABPs and {OBPs}, each with $N=800$ particles on a two-dimensional domain of length $L_x=80$ and width $L_y=50$. Particles interact with a soft repulsive ``harmonic potential," such that $\U(\br)=\frac{\varepsilon }{2} \left[1 - |\br|\right]^2$ when the interparticle separation $|\br|$ is less than 1, and is 0 otherwise. Here $\varepsilon$ is the interaction strength and is equal to 50 for the interacting simulations and 0 otherwise. The activity landscape experienced by ABPs, $v(x)$, linearly interpolates from $v_h=20$ to $v_\ell=1$ over an interval $\Delta x_1=2.5$ and back to $v_h$ over an interval $\Delta x_2=12.5.$ The temperature landscape experienced by {OBPs}, $T(x)$, is the same. We used a time step dt$=2.5\times10^{-4}$, and ran the simulation until $t_{\rm f}=10^6$. The density profiles are averages over $5 \times 10^5$ evenly-spaced snapshots taken throughout the simulations.

Fig.~\ref{fig:EOSpassive} shows simulations of {2d OBPs}, interacting with the harmonic potential, with the parameters dt$=5\times 10^{-4}$, $t_{\rm f}=10^5$, $N=400$, $L_x=50$, $L_y=15$, $\varepsilon=50$, $T(x) \in [1,10]$, $\Delta x_1=8$, and $\Delta x_2=24$. Cubic interpolation is used for $T(x)$. The {entropy production rate} and stress profiles were measured at $2 \times 10^7$ and $2 \times 10^6$ evenly-spaced intervals, respectively.

Fig.~\ref{fig:momentum-source-abp} shows simulations of {2d} ABPs, interacting with the harmonic potential, with the parameters dt$=5\times 10^{-4}$, $t_{\rm f}=4\times 10^5$, $N=800$, $L_x=80$, $L_y=15$, $\varepsilon=50$, $v(x) \in [1,20]$, $\Delta x_1=10$, and $\Delta x_2=30$. Cubic interpolation is used for $v(x)$. Data are averaged over $n_{\rm seed}=260$ seeds, and profiles are averaged over $2 \times 10^5$ snapshots per seed.

Fig.~\ref{fig:rtp-u-v-exact-sol}c shows the current directly measured in simulations of noninteracting {1d} RTPs with parameters dt$=10^{-3}$, $t_{\rm f}=10^5$, and $L_x=30$. The activity landscape is given by the sinusoidal profile $v(x)=2 \left[0.4 \sin\big(\frac{2\pi x}{L_x}\big) + 0.2 \sin\big(\frac{4\pi x}{L_x}\big)\right]$. The particles experience an external potential $U(x)=\varepsilon \sin\big(\frac{4\pi x}{L_x}\big)$. Each point results from averaging over 1000 single-particle simulations.

Fig.~\ref{fig:mean-field-schematic} shows simulations of interacting {1d} RTPs with dt$=5\times 10^{-4}$, $L_x=10$, and sinusoidal activity profile $v(x)= 20\left[0.5 \sin\big(\frac{2\pi x}{L_x}\big) + 0.4 \sin\big(\frac{4\pi x}{L_x}\big)\right]$. The RTPs interact via an ``inverse harmonic potential" $\U(x)=\frac{\varepsilon}{2} \big[1-x^2\big],$ with varying $\varepsilon$. The simulation lengths $t_{\rm f}$ and number of seeds $n_{\rm seed}$ vary with the number of particles $N$ as indicated in the following list of tuples $(N,t_{\rm f},n_{\rm seed})$: $(12,6\times 10^7,480)$, $(25,6\times 10^7,240)$, $(50, 5\times 10^7,400)$, $(100,8\times 10^6,240)$, $(200,4\times 10^6,360)$, and $(400,2\times 10^6,400)$.

{
\vspace{0.1in}
\noindent\textit{Derivation of $\sigma_\subIK$.} 
Here we provide a derivation of the Irving-Kirkwood stress tensor $\bsigma_\subIK$~\cite{irving1950statistical}, which we use throughout the text. Any  conservative force $\bF_{ij} = -\nabla \U(\br_i-\br_j)$ is reciprocal and thus satisfies}
\begin{align}
 {\sum_{ij} \bF_{ij}\delta(\br -\br_i)}
 &{= \frac{1}{2} \sum_{ij} (\bF_{ij} - \bF_{ji}) \delta(\br -\br_i)}\\
 &{= \frac{1}{2} \sum_{ij} \bF_{ij}\big[\delta(\br -\br_i) - \delta(\br -\br_j) \big]\;,}\label{eq:Irving-Kirkwood}
 \end{align}
{ where the second equality stems from reindexing. Introducing $\br_{ij} \equiv \br_i - \br_j$, the chain rule shows that}
 \begin{equation}
     {\partial_\lambda \delta(\br - \br_j - \lambda\br_{ij})= - \br_{ij} \cdot  \nabla \delta(\br - \br_j - \lambda\br_{ij}) \;,}
 \end{equation}
{so that}
\begin{equation}
    {\delta(\br -\br_i) - \delta(\br -\br_j)
    = -\nabla \cdot \bigg[\br_{ij} \int_0^1 d\lambda\ \delta(\br - \br_j - \lambda\br_{ij}) \bigg]\,.}
\end{equation}
{In turn, this allows rewriting Eq.~\eqref{eq:Irving-Kirkwood} as}
\begin{align}
{ \sum_{ij} \bF_{ij}\delta(\br -\br_i)}
 &{=   - \nabla \cdot \sum_{ij} \frac{\br_{ij}}2 \otimes \bF_{ij} \int_0^1 d\lambda\ \delta(\br - \br_j - \lambda \br_{ij}) }\\
 &{\equiv \nabla\cdot \bsigma_\subIK\,,}
\end{align}
{where the last line is the definition of $\bsigma_\subIK$.}

\bibliography{refs-resub}

\end{document}